\documentclass[aps,prl,amssymb,showpacs,amsmath,superscriptaddress,preprintnumbers,reprint]{revtex4-1}
\pdfoutput=1 %This is for the ArXiv submission only

\usepackage{natbib}
\usepackage{soul}
\usepackage{graphicx}
\usepackage{tikz}
\usetikzlibrary{snakes}
\usepackage{pdfpages}

\makeatletter
\AtBeginDocument{\let\LS@rot\@undefined}
\makeatother

\usepackage{color}

\renewcommand{\epsilon}{\varepsilon}

\begin{document}

\title{Multiple singularities of the equilibrium free energy in a one-dimensional model of soft rods}

\author{Sushant Saryal}
\affiliation{Indian Institute of Science Research and Education, Pashan, Pune, India.}
\author{Juliane U. Klamser}
\affiliation{
Laboratoire de Physique Statistique, D\'{e}partement de physique de l'ENS, 
Ecole Normale Sup\'{e}rieure, PSL Research University, Universit\'{e} Paris Diderot, 
Sorbonne Paris Cit\'{e}, Sorbonne Universit\'{e}s, UPMC Univ. Paris 06, CNRS, 75005 
Paris, France.}
\author{Tridib Sadhu}
\affiliation{Tata Institute of Fundamental Research, Mumbai 400005, India.}
\affiliation{Coll\`{e}ge de France, 11 place Marcelin Berthelot, 75231 Paris Cedex 05, France.}
\author{Deepak Dhar}
\affiliation{Indian Institute of Science Research and Education, Pashan, Pune, India.}

\begin{abstract}
There is a  misconception, widely shared amongst physicists, that the equilibrium free energy of a one-dimensional classical model with strictly finite-ranged interactions, and  at non-zero temperatures, can not show any singularities  as a function of the coupling constants. In this Letter, we discuss an instructive  counter-example.  We consider thin rigid linear rods of equal length $2 \ell$ whose centers lie on a one-dimensional lattice, of lattice spacing $a$. The interaction between rods is a soft-core interaction, having a finite energy $U$ per overlap of rods.  We show that the equilibrium  free energy per rod $\mathcal{F}(\tfrac{\ell}{a}, \beta)$, at inverse temperature $\beta$, has an infinite number of singularities, as a function of $\tfrac{\ell}{a}$. 
\end{abstract}


\maketitle

There is a common belief amongst physicists that in any one-dimensional (1-d) classical system, in thermal equilibrium, having  strictly finite-ranged pairwise interactions,  the thermodynamic potential cannot show  a singular dependence on the control parameters \cite{Minlos1975}.
The origin of this folk wisdom  is perhaps an  unsubstantiated generalization of 
a rigorous result due to van Hove \cite{VANHOVE1950} on the absence of phase 
transitions in a one-dimensional system of particles with a non-vanishing 
hard-core length and finite-ranged inter-particle interaction. This result was 
later extended to lattice models \cite{RUELLE1999} and long-ranged interactions 
having a  power-law decay with distance \cite{RUELLE1968,DYSON1969,FROLICH1982}. 
The belief further grew out of essentially two (correct) arguments: one, about 
the absence of phase transitions as a function of temperature in 1-d  models 
having a finite-dimensional irreducible transfer matrix and second, the Landau 
argument about the absence of symmetry-breaking in 1-d systems, when creating a 
domain-wall has a finite energy cost \cite{Landau}. Several counter-examples of 
equilibrium phase transitions in 1-d models have been known for a long time: DNA 
unzipping \cite{KITTEL1969,DAUXOIS1993}, interface depinning \cite{CHUI1981}, 
and condensation in zero-range models \cite{GROSSKINSKY2003}.  But, the 
incorrect belief persists. A necessary and sufficient condition for the 
existence of phase transitions in 1-d systems is hard to formulate.  This 
question was discussed in some detail recently by Cuesta and Sanchez 
\cite{CUESTA2004},  who  provided a sharper criteria for the absence of phase 
transitions, based on a generalized Perron-Frobenius-Jentzsch theorem. The 
general understanding is that singularities in the free energy come from the 
degeneracy of the largest eigenvalue of the transfer matrix which can occur when 
the conditions required for the Perron-Frobenius-Jentzsch theorem to hold are 
not met.

In this Letter, we discuss an example of a  1-d system that undergoes an {\em infinite} number of phase transitions, even though the largest eigenvalue remains non-degenerate. The singularities are robust, geometrical in origin, and come from the changes in the structure of the interaction Hamiltonian as a function of the separation between particles. This is a simple,  instructive  example, and  it uses a different mechanism of generating singularities in the thermodynamic functions than the earlier models studied.

\begin{figure}[t]
\includegraphics[scale=0.40]{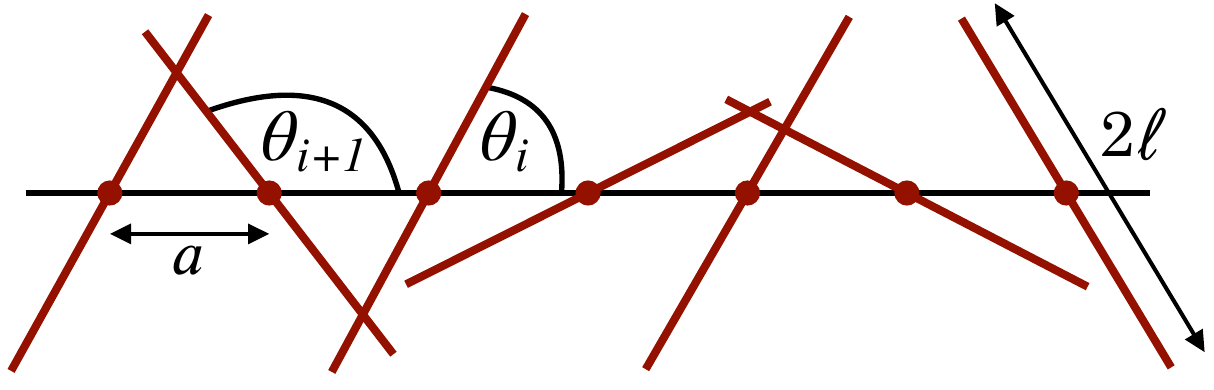}
\caption{A configuration of $7$ rods on a line. Here, $a$ is the spacing between rods.  In the displayed configuration, the number of nearest neighbor overlaps $n_1 =3$ and the number of next nearest overlaps $n_2 =1$. \label{fig:rods}}
\end{figure}

In its simplest version, the model consists of soft linear rigid rods of equal length $2 \ell$,  whose midpoints are  fixed at  the lattice sites of a 1-d lattice of lattice spacing $a$.  The rods are free to rotate in the plane, as illustrated in figure \ref{fig:rods}, where a configuration of $N$ rods is specified by a set of $N$ angles $\theta_i$, with $0 \leq \theta_i \leq \pi$, for $i = 1$ to $N$. We assume that there is an interaction  between the rods, which depends on their overlap. Each overlap between a pair of nearest neighbor rods costs a constant energy $U_1$; between a pair of next nearest neighbors the overlap energy is $U_2$, and so on. Let $n_r$ be the number of  pairs of the $r$-th neighbor rods that overlap (see figure \ref{fig:rods}). Clearly, $n_r$ is zero, if $r > \tfrac{2\ell}{a}$. The total energy of the system is 
\begin{equation}
\mathcal{H} = \sum_{i} n_{i} U_{i}\,.
\end{equation}
This is similar to the hard-rod model that has been studied a lot in the literature, starting with Onsager \cite{ONSAGER1949,casey1969, kantor2009,gurin2011}. It differs in two significant ways: the centers of the rods are fixed on a lattice, and we allow $U_i$ to be any sign (attractive or repulsive  soft-cores). A somewhat similar model  of non-spherical molecules whose centers are fixed at equi-spaced points along a line, but orientations can change,  was studied in \cite{casey1969}. 

Let $\mathcal{F}(\tfrac{\ell}{a} = \kappa, \beta)$ denote the free energy per rod of this system, in equilibrium, at inverse temperature $\beta$.  We will show that $\mathcal{F}(\kappa, \beta)$ is an analytic function of $\beta$, as expected, but has  a non-analytic dependence on $\kappa$. In fact, there are infinitely many transitions: as $\kappa$ is varied, $\mathcal{F}(\kappa, \beta)$ is singular at every positive integer values of $\kappa$,  for all  $\beta$. The singularities remain unchanged irrespective of the sign of $U_i$, whether the interaction is repulsive or attractive.  We will show that there are also other singularities at some non-integer values of $\kappa$. For example, the  probability distribution of orientations changes  qualitively when $\kappa$ is changed across $\tfrac{1}{\sqrt{2}}$.

For simplicity of presentation, we begin with the simple case: $U_1= \infty$. This is the case of hard-rods, where no nearest-neighbor overlaps are allowed, thus $n_i = 0$ for all $i \ge 1$. Then, without loss of generality, we may assume $U_i=0 $ for all $i\ge 2$, which corresponds to only nearest neighbor  hard-core interactions. In this case, let $\mathcal{F}_1(\kappa)$ denote the free energy per site in the thermodynamic limit (due to hard-core interactions $\beta$ is irrelevant and hence omitted). Then, using the transfer matrix technique, $\mathcal{F}_1(\kappa) = -\log \Lambda(\kappa)$, where $\Lambda(\kappa)$ is the largest eigenvalue of the integral equation

\begin{equation}
\Lambda(\kappa) \psi_\kappa(\theta) = \int_{0}^{\pi} \frac{d\theta'}{\pi} ~~ T_\kappa(\theta, \theta') \psi_\kappa(\theta')\,,
\label{eq:eigenvalue}
\end{equation}
with $\psi_\kappa(\theta)$ being the associated eigenvector. The transfer matrix $T_\kappa(\theta',\theta)$ has matrix elements $0$ or $1$ depending on whether a pair of nearest neighbor rods with angles $(\theta',\theta)$ overlap or not. 

We will show below that  this system shows three types of singularities: (i) $ {\mathcal F}''_1(\kappa)$ is discontinuous at $\kappa = \tfrac{1}{2}$, (ii) for $\kappa$ near $1$, say $\kappa = 1 +\epsilon$, with $|\epsilon| \ll 1$,  ${\mathcal F}'_1(\kappa)$ diverges as  $ \log (|\epsilon|)$, and  (iii) for $ \tfrac{1}{\sqrt{2}} < \kappa < 1$, the probability distribution of orientations $P_\kappa(\theta)$ has square-root singularities as a function of $\theta$, which are not present for lower values of $\kappa$. 

\begin{figure}
\includegraphics[scale=0.65]{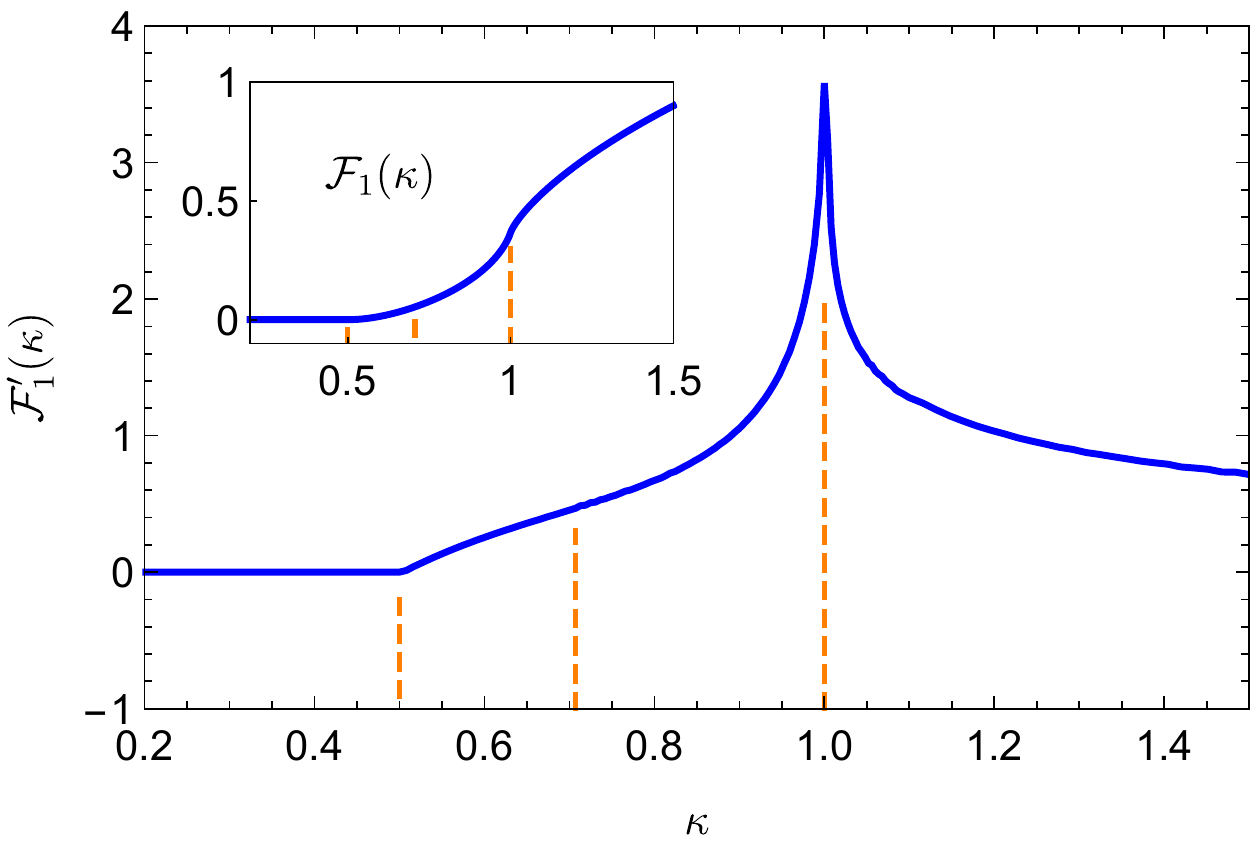}
\caption{First derivative of the free energy $\mathcal{F}'_1(\kappa)$ for hard-core nearest neighbor interaction between rods ($U_1 =\infty$). The inset shows the monotonic increase of $\mathcal{F}_1(\kappa)$ as a function of $\kappa$.} \label{fig:derivative}
\end{figure}
\begin{figure}
\includegraphics[scale=0.8]{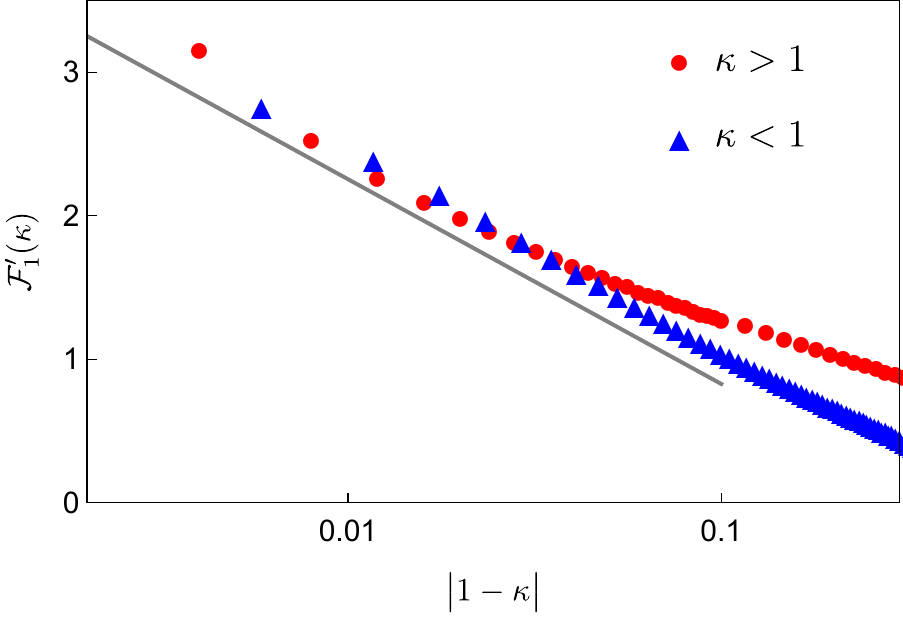}
\caption{Logarithmic divergence of the first derivative of the free energy $\mathcal{F}'_1(\kappa)$ near $\kappa=1$, for $U_1 = \infty$. \label{fig:logsing}}
\end{figure}

The numerical verification of these analytical results is  shown in figures $2$-$4$, obtained by numerically diagonalizing the transfer matrix, using $1000$ grid points for the integartion range of $\theta = [0,\pi]$.  In figure \ref{fig:derivative}, ${\mathcal F'}_1(\kappa)$ is exactly zero for $\kappa < \tfrac{1}{2}$, and nonzero for $\kappa > \tfrac{1}{2}$, initially increasing linearly. Near $\kappa =1$, it has a sharp peak. In figure \ref{fig:logsing}, ${\mathcal F'}_1(\kappa)$ shows a nearly linear dependence on $\log |\kappa -1|$. 

\begin{figure}
\includegraphics[width=0.45\textwidth]{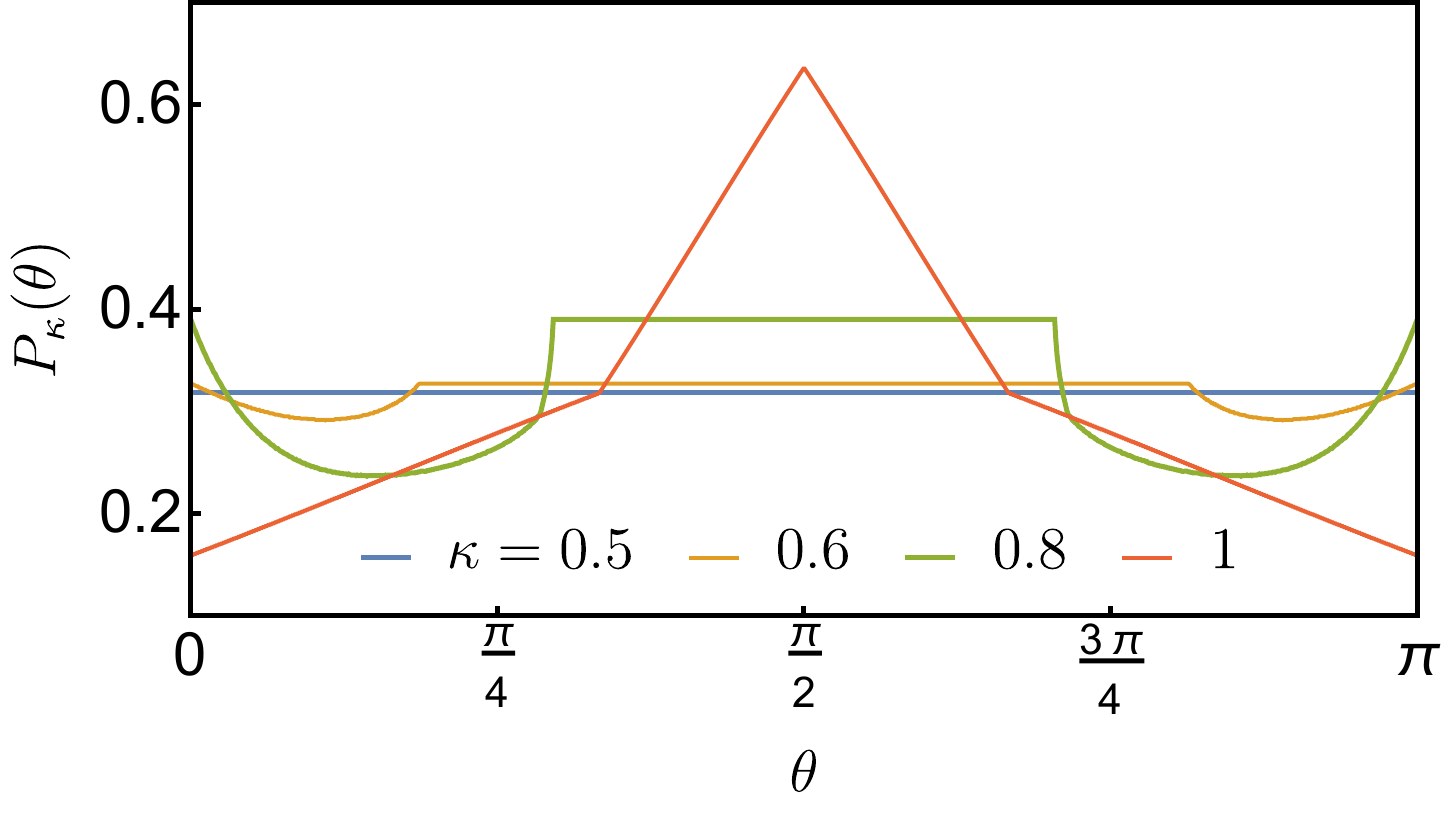}
\caption{Probability distribution of the orientation of the rods generated from the eigenvector $\psi_\kappa(\theta)$ associated to the largest eigenvalue of the transfer matrix.\label{fig:prob}}
\end{figure}

We determine the probability distribution of orientations $P_\kappa(\theta)$ from the eigenvector $\psi_\kappa(\theta)$ of the transfer matrix.  This is plotted in figure \ref{fig:prob}. For $\kappa < \tfrac{1}{2}$, all angles are equally likely, and $P_\kappa(\theta)$ takes a constant value $\pi^{-1}$.  For $\tfrac{1}{2} < \kappa < \frac{1}{\sqrt{2}}$, $P_\kappa(\theta)$ has a non-trivial dependence on $\theta$ when $\vert\cos \theta\, \vert > \tfrac{1}{2\kappa }$, but the derivative $P'_\kappa(\theta)$ remains finite. In the range $\frac{1}{\sqrt{2}} < \kappa <1$, $P_\kappa(\theta)$ has a square-root cusp singularity, when  $\sin \theta = \kappa$. 
There is no clear signature of this singularity  in the functional dependence of ${\mathcal F}_1(\kappa)$ on $\kappa$.

%The eigenvector can be determined similarly; the probability $P(\theta)=\psi(\theta)^2$ obtained this way is in figure \ref{fig:prob}. Two two special values of $\kappa$ can been seen: below $\kappa=1/2$ the probability distribution is uniform and as $\kappa$ crosses $1/2$ the distribution develops two singularities with discontinuous derivative. A third singularity develops at $\kappa=1$. 

\begin{figure}
\includegraphics[scale=0.37]{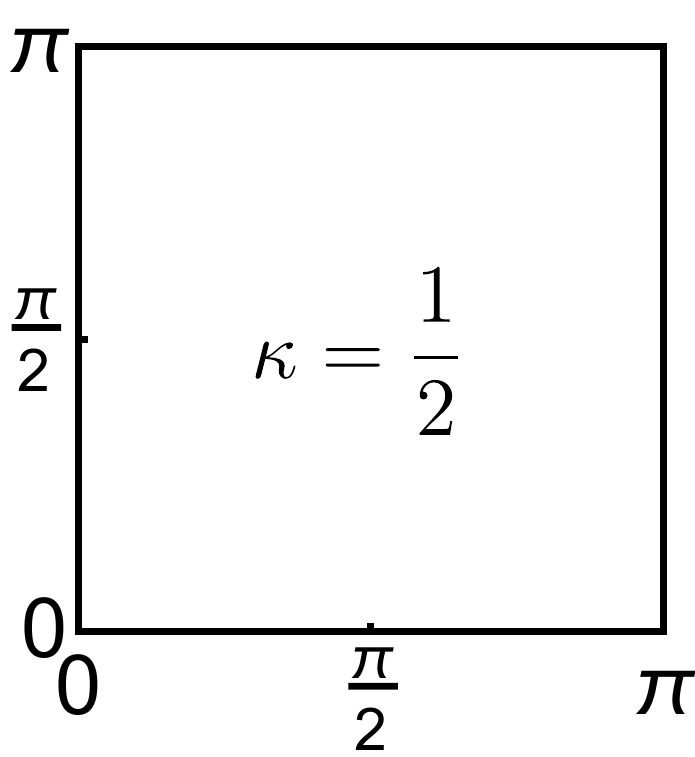}
\includegraphics[scale=0.37]{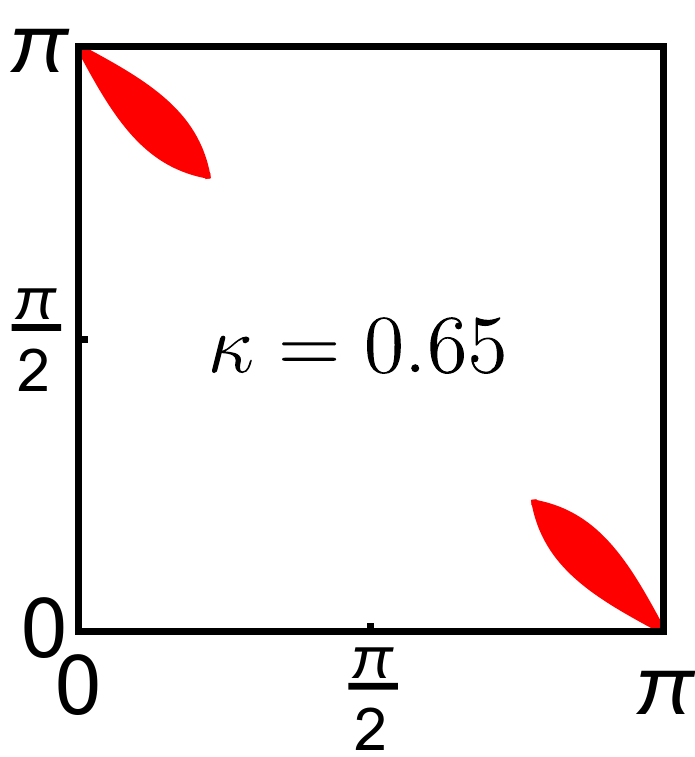}
\includegraphics[scale=0.37]{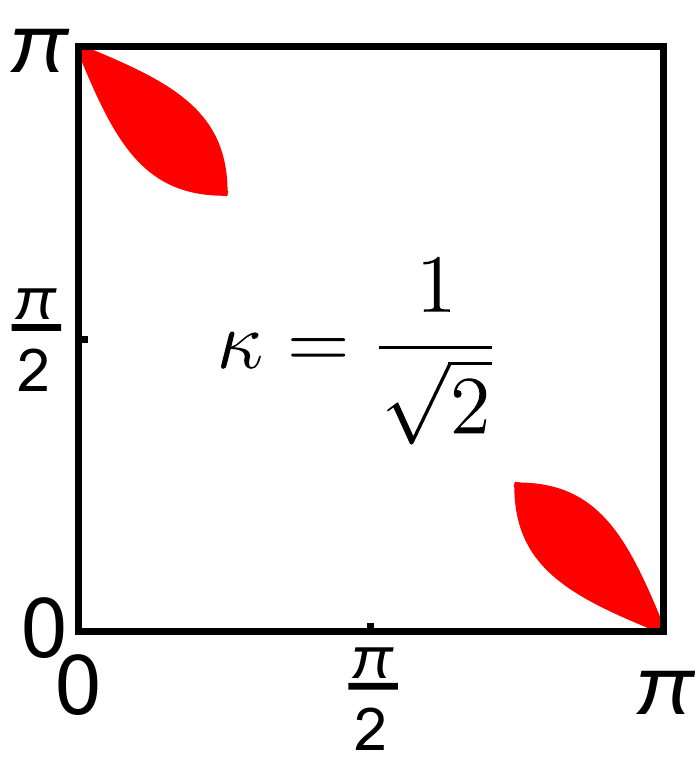}
\includegraphics[scale=0.37]{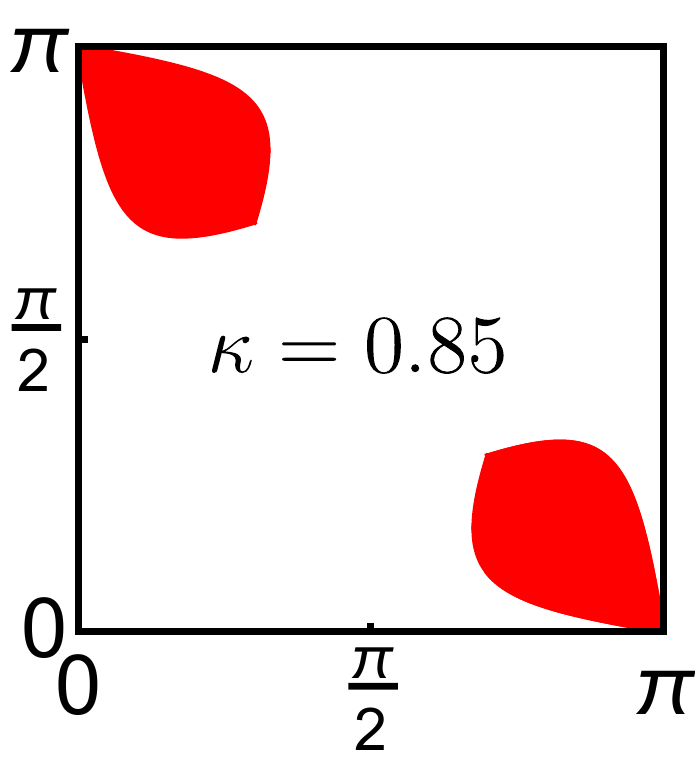}
\includegraphics[scale=0.37]{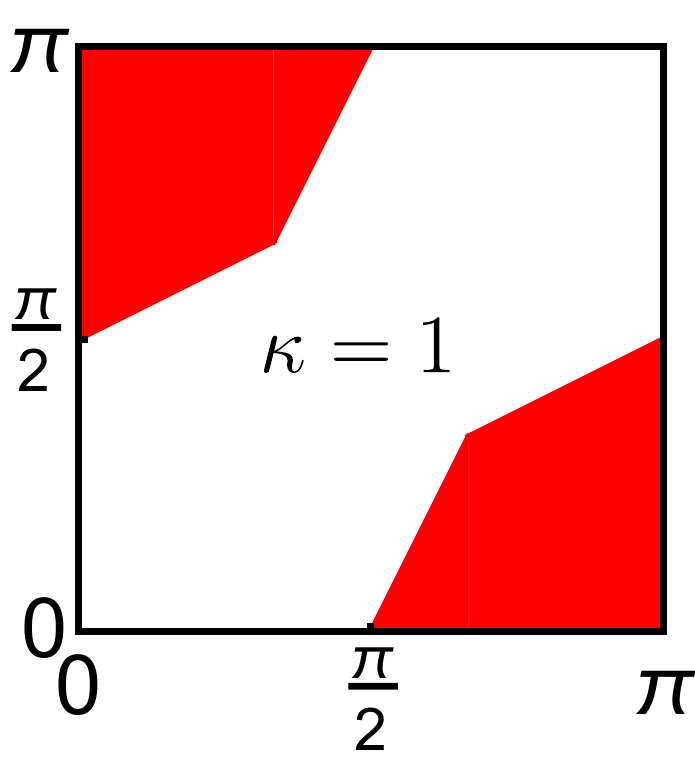}
\includegraphics[scale=0.37]{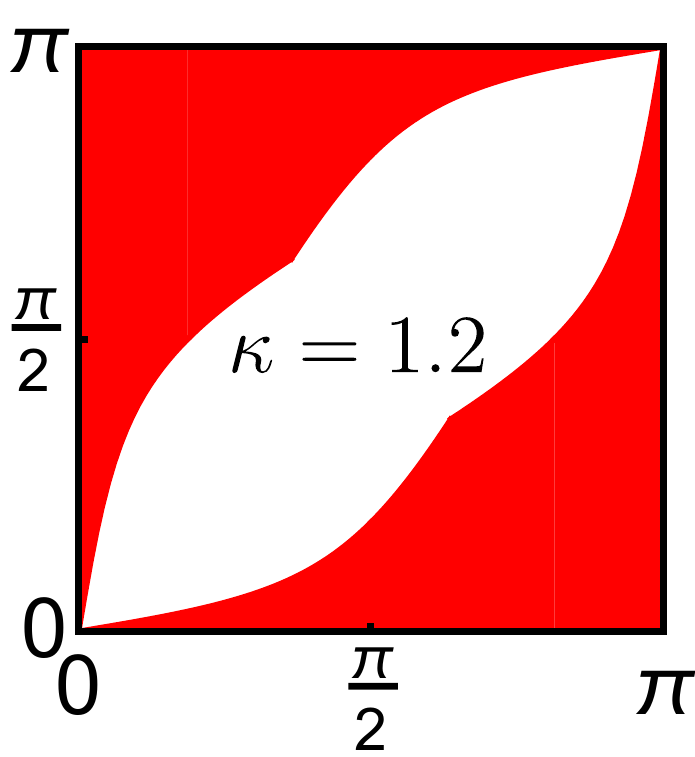}
\caption{The transfer matrix $T_\kappa(\theta',\theta)$ on the $\theta$-$\theta'$ plane, for different values of $\kappa$. The shaded regions denote $(\theta,\theta')$ values where the rods overlap, and  $T_\kappa=0$. In the plain regions rods do not overlap and $T_\kappa=1$.\label{fig:Tmatrix}}
\end{figure}

The source of these singularities is geometric in nature,  and can be seen most simply in the structure of the transfer matrix. This is illustrated figure \ref{fig:Tmatrix}. Here the shaded regions
in the $\theta$-$\theta'$ plane correspond to values of $(\theta,\theta')$ where the rods intersect, and the matrix element is $0$, whereas the plain regions correspond to non-intersecting rods, and the matrix element is $1$. The equation of the boundary of the shaded region is easily written down from simple geometry (see supplementary material for details).  As $\kappa$ is increased, the shaded regions grow in size, and the eigenvalue of the transfer matrix decreases. For  $\tfrac{1}{\sqrt{2}} < \kappa < 1$, the slope of the boundary of the shaded region becomes infinite or zero at some points. When $\kappa=1$, the boundary becomes a set of straight lines. For $\kappa >1$, the two shaded patches, which are disjoint when $\kappa < 1$, merge into a single connected shaded region. We will show that precisely these topological changes in the structure of the available phase space lead to the singularities in the free energy function ${\mathcal F}_1(\kappa)$.

Let us first discuss the singularity at $\kappa =\tfrac{1}{2}$.  For $\kappa<\tfrac{1}{2}$, no overlap is possible, and the rods can orient freely without any cost of energy. The associated transfer matrix $T_\kappa(\theta',\theta)=1$ for all angles, and there are no shaded regions. The largest eigenvalue is $\Lambda(\kappa)=1$ and the corresponding eigenvector $\psi_\kappa(\theta) =$ constant. As $\kappa$ is increased beyond $\tfrac{1}{2}$ the nearest neighbor interaction sets in. If we define $\kappa = \tfrac{1}{2} + \epsilon$, then it is easily seen that for small $\epsilon >0$, the area of the shaded regions in the $\theta$-$\theta'$ plane grows as $\epsilon^2$.   Then, treating the shaded regions as perturbation, the first order perturbation theory  immediately gives 
\begin{equation}
\Lambda(1/2 +\epsilon) = 1 - C \,\epsilon^2  + {\rm ~higher~order~in~}\epsilon.
\end{equation}
We find that the constant $C = \tfrac{32}{3 \pi^2}$ (details in the supplementary material). Thus, at $\kappa = \tfrac{1}{2}$, the second derivative of the free energy $\mathcal{F}''_1(\kappa)$ with respect to $\kappa$ is discontinuous.

We now discuss the singularity at $\kappa =1$. For this value, the boundary of the excluded region in the $\theta$-$\theta'$ plane becomes a set of straight lines (see figure \ref{fig:Tmatrix}). Then, the transfer matrix $T_\kappa(\theta',\theta)$ can be exactly diagonalized by converting the integral eigenvalue equation \eqref{eq:eigenvalue} into a second order differential equation. The details are given in the supplementary material. We find that the largest eigenvalue of the transfer matrix for $\kappa =1$ is given by $\Lambda(1) = \left[ 3 \sqrt{2}\arcsin(\tfrac{1}{3})\right]^{-1}$.

\begin{figure}
\includegraphics[scale=0.5]{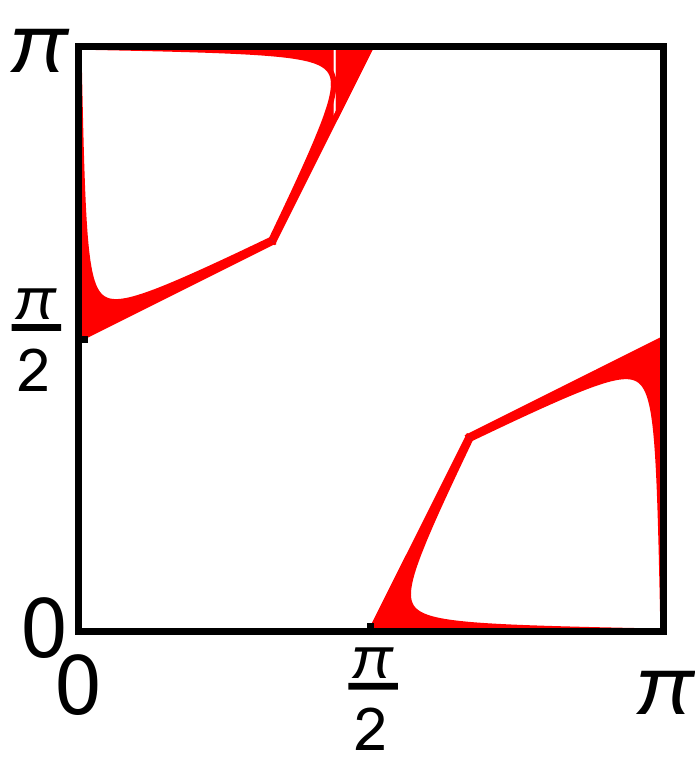}
\caption{The picture shows the matrix $\Delta T=T_{1-\epsilon}-T_1$, for $\epsilon=0.02$ on the $\theta$-$\theta'$ plane.  In the shaded region $\Delta T=1$, whereas in the plain region it is $0$. The area of the shaded region varies as $\epsilon \log \tfrac{1}{\epsilon}$, for small $\epsilon$. }  \label{fig:DeltaT}
\end{figure}

For $\kappa$ near $1$, if we write $\kappa = 1 - \epsilon$ and define $\Delta T = T_{ 1 - \epsilon} -T_1$, then, to the first-order in $\epsilon$,  the change in the eigenvalue $\Lambda(\kappa)$ equals $\langle \psi_1|\Delta T|\psi_1\rangle$, where $\psi_1(\theta)$ is the eigenvector of the transfer matrix corresponding to the largest eigenvalue at $\kappa =1$.  This change is shown in figure \ref{fig:DeltaT}.  The curved boundary of the disallowed region near $(\theta,\theta') \equiv (0,\tfrac{\pi}{2})$ tends to a hyperbola, and as $\epsilon$ tends to zero, the area of the the shaded region in figure \ref{fig:DeltaT} tends to zero, but only as $\epsilon \log \tfrac{1}{\epsilon}$. Moreover, the eigenvector $\psi_1(\theta)$ is positive everywhere, with the ratio between its maximium and minimum values remaining finite. This implies that the change in the matrix element has the same qualitative dependence on $\epsilon$ as the area of the shaded regions. Therefore, we conclude that 
\begin{equation}
\Lambda(1 - \epsilon) = \Lambda(1) + K_1 \epsilon \log \frac{1}{\epsilon} + K_2 \epsilon +{\rm ~higher ~order  ~terms},
\end{equation}
where $K_1$ and $K_2$ are positive constants.  A similar argument holds for negative $\epsilon$ and the details are given in the supplementary material. 

We now discuss the singularity at $\kappa = \tfrac{1}{\sqrt{2}}$. For this we consider the range $\tfrac{1}{\sqrt{2}} < \kappa < 1$, and define $\theta_0 = \sin^{-1} \kappa$. Then, as long as the angle of a rod $\theta \in [\theta_0,\pi - \theta_0]$, it can be easily seen, that there is no overlap with its neighbor for any angle $\theta'$ of the latter. On the other hand, if $\theta$ is outside this interval, the rods can intersect, if $\theta'$ lies in the intervals $[ \phi_1, \phi_2]$ and $[ \pi-\phi_2, \pi-\phi_1]$, with the expression for $\phi_1$ and $\phi_2$ given in the supplementary material. The important point is that the length of the intervals $\vert \phi_2 - \phi_1\vert$ varies as $\sqrt{\theta_0 - \theta}$ for $\theta\rightarrow\theta_0$.   Then, from the eigenvalue equation \ref{eq:eigenvalue}, we see that
\begin{equation}
\psi_\kappa(\theta) = K_3 -  K_4\int_{\phi_1(\theta)}^{\phi_2(\theta)} \psi_\kappa(\theta') d\theta',
\end{equation}
where $K_3$ and $K_4$ are functions of $\kappa$ only. Using this fact that $\psi_\kappa(\theta')$ is bounded by non-zero constants, both from above and below, we see that, for $\theta$ approaching $\theta _0$ from below
\begin{equation}
\psi_\kappa(\theta) \approx K_3 - K_5 \sqrt{ \theta_0 -\theta},
\end{equation}
where $K_5$ depends only on $\kappa$.  This shows that $\psi_\kappa(\theta)$ has a cusp singularity at $\theta = \theta_0$. As the probability density $P_\kappa(\theta)$ is proportional to $\psi_\kappa(\theta)^2$, it also has a cusp singularity for $ \theta = \arcsin \kappa$.

Our above arguments can be readily generalized to the case of soft rods ($U_1 \neq +\infty$), but keeping $U_i = 0$ for $i > 1$. The matrix $\Delta T$ only gets multiplied by a factor $(1 - e^{ -\beta U_1})$. In fact, one can even determine the exact eigenvalues of the transfer matrix at $\kappa =1$, for an arbitrary pair-potential $U_1$. This is given by (see supplementary material)
\begin{equation}
\Lambda(1) = \frac{(1 - e^{-\beta U_1})}{3 \sqrt{2}} \left[ \arctan \frac{(1 - e^{-\beta U_1})}{\sqrt{2}(2 + e^{-\beta U_1})}\right]^{-1}\,.
\end{equation}

For soft pairwise interactions, overlaps between pairs of rods beyond the nearest neighbors are allowed. In the case, where such overlaps cost a non-zero amount of energy, \textit{i.e} $U_i \ne 0 $ for $i > 1$, one can treat these pair-interactions $U_i$, as perturbations to the problem with only non-zero $U_1$. Noting that the overlap region in the $(\theta_j, \theta_{j + i})$-plane, for $i > 1$,  again has a similar hyperbolic shape,  we  see that at all integer values of $\kappa = i$ the largest eigenvalue $\Lambda(\kappa)$ has singularities of the form $ U_i (i - \kappa) \log \vert\tfrac{1}{\kappa - i}\vert$.

\begin{figure}
\includegraphics[width=0.48\textwidth]{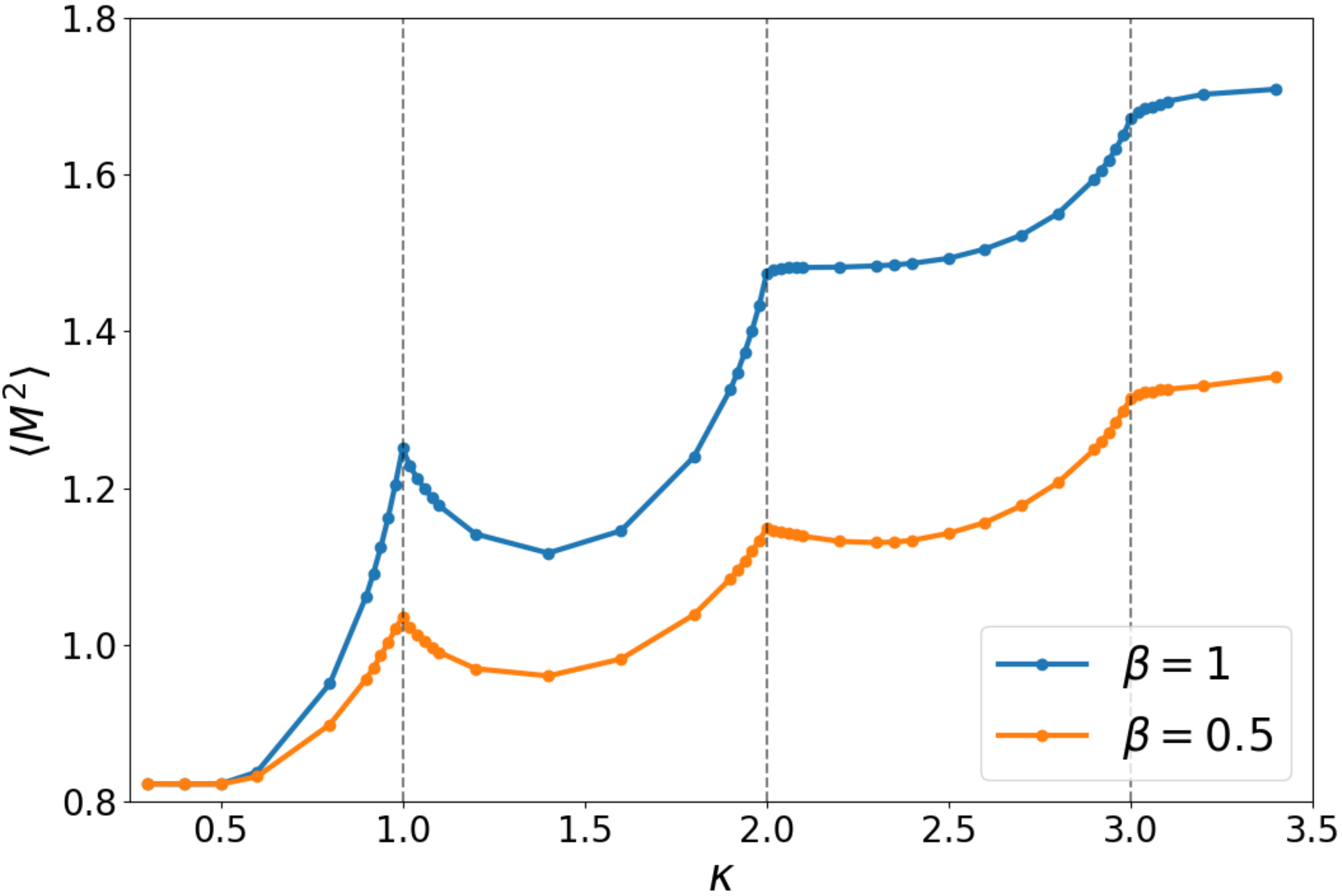}
\caption{Variance of the angular distribution of rods generated from Monte Carlo simulations of a system of $100$ rods and averaged over $10^6$ sample configurations. \label{fig:variance}}
\end{figure}

In figure \ref{fig:variance}, we present evidence of these additional transitions from Monte Carlo simulations. We took $U_i =1$ for all $i$. Clearly, we have no long-range correlations in the system, and  $\langle \theta \rangle=\tfrac{\pi}{2}$, for all $\kappa$.  A signature of the transitions can be seen in the variance of the angle defined by $\langle M^2\rangle=\tfrac{1}{N}\langle \left[\sum_i (\theta_i-\tfrac{\pi}{2})\right]^2\rangle$. The variance clearly shows a singularity at all integer values of $\kappa$. Also, the positions of the singularities do not depend on the value of $\beta$, as long as it remains nonzero.

The reason why the conditions for the applicability of the van Hove theorem are not met is quite clear. As the van Hove theorem demands, the matrix elements are analytic functions of $\beta$; however, in our case they are non-analytic (in fact discontinuous) functions of the control parameter $\kappa$. This non-analyticity is generic to all hard-core (or soft-core) models, and is at the root of the singular behavior found in the problem discussed here. Note that analyticity of the interaction potential as function of distance is not required for a well-behaved thermodynamic limit.

We note that the free energy $F(\kappa,\beta)$ is a non-convex function of $\kappa$ (see inset of figure \ref{fig:derivative}). Here, $\kappa$ is a parameter that specifies the number of rods per unit length in the system, and convexity of the free energy as a function of density is a fundamental property, which is essential for thermodynamic stability. In our model,  the spacing between particles is fixed and can not be changed. Hence a convex envelope construction, \textit{\`a la} Maxwell,  is not possible, and convexity is not assured. In fact, if the spacing between rods is allowed to vary, then the free energy has no singularities,  in agreement with all the previous studies of this model \cite{casey1969, kantor2009,gurin2011}. 

Additionally, we note that in our system, for all finite $\kappa$, the correlation length remains finite, and the largest eigenvector remains non-degenerate. Moreover, the behavior of the free energy here is different from the familiar first order phase transitions, where the correlation length remains finite at the transition point, and the first derivative of the free-energy is discontinuous. In our case, the first derivative is {\it divergent}  at the transition points.

	Are the points of non-analyticity of the free energy in our system also {\it phase transition points} between distinct phases, or are they similar to the fluid-fluid transition (\textit{e.g.} the liquid-gas transition), where a non-analyticity in the free energy occurs along a line within the same fluid phase? To answer this question, we consider a particular observable quantity in the equilibrium state: the fraction of $i$-th neighbor rods that overlap, as an order parameter, which is proportional to $\frac{\partial {\mathcal F}}{\partial U_i}$. This is exactly zero for $\kappa \leq i$, and non-zero otherwise. This shows that distinct values of $\lfloor \kappa \rfloor$ ($\lfloor \cdot \rfloor$ denotes floor function) correspond to thermodynamically  distinguishable distinct phases of the system.
Of course, these phases could be further split using additional criteria, \textit{e.g.} by the behavior of the distribution of angles.

It is easy to construct other models which show similar behavior. 
For example, consider a chain of Ising spins $\sigma_i$,  placed on a lattice of uniform  spacing $a$. The Hamiltonian of the system is  $H= -\sum_{(i,j)} J(r_{ij})\sigma_i\sigma_j$, where $J(r)$ is  a distance-dependent exchange interaction $J(r)$,  and $r_{ij}$ is the distance between the sites $i$ and $j$.  If we  choose, $J(r) = 1-r$, for $0 < r <1$, and zero for $r >1$, there is no long-range order in the problem. However, as the lattice spacing $a$ is varied, the free energy becomes a non-analytic function of $a$, at all integer values of $\tfrac{1}{a}$, following the same reasoning as in our model.

In summary, we have discussed a  mechanism of phase transitions,  which is simple, but has not been sufficiently emphasized in the past. We have illustrated this mechanism with the example of a  model of soft rods on a lattice in 1-d with short range interactions, which shows an infinite number of phase transtitions. The model differes from the well-studied models of the past only in the aspect that the centers of rods are placed on a regular lattice, and the distance between them cannot change, except as a global parameter.  One would expect similar behavior to occur for objects of different shapes, like crosses, or T- or Y-shapes. The singularities will also occur in higher dimensions. We have studied the system of soft rods in 2-dimensions, which  shows similar phase transitions, at $\tfrac{1}{a} = \sqrt{m^2 + n^2}$, where $m$ and $n$ are any integers.  These will be reported in a future publication \cite{Klamser2018}.

\bibliography{citation}

\clearpage
\newpage

\begin{widetext}
	\begin{center}
		%\includepdf[pages={1-8},\pagecommand=\section{Supplemental Material}]{supplement.pdf}
		

\section*{Supplementary material}

# Supplemental material for: Multiple singularities of the equilibrium free energy in a one-dimensional model of soft rods

Sushant Saryal,[1] Juliane U. Klamser,[2] Tridib Sadhu,[3,4] and Deepak Dhar[1]

[1]*Indian Institute of Science Research and Education, Pashan, Pune, India.*
[2]*Laboratoire de Physique Statistique, Département de physique de l'ENS,*
*Ecole Normale Supérieure, PSL Research University,*
*Université Paris Diderot, Sorbonne Paris Cité, Sorbonne Universités,*
*UPMC Univ. Paris 06, CNRS, 75005 Paris, France.*
[3]*Tata Institute of Fundamental Research, Mumbai 400005, India.*
[4]*Collège de France, 11 place Marcelin Berthelot, 75231 Paris Cedex 05, France.*

We present some of the algebraic details of derivations, and additional results from Monte Carlo simulations. To be specific, we give detailed expressions of the overlap region, an analysis of the singularities, and an exact diagonalization of the transfer matrix. The results from Monte Carlo simulations are about the probability distribution of the orientation of a rod at different values of $\kappa$.

## I. THE STRUCTURE OF THE TRANSFER MATRIX

We discuss the transfer matrix $T_\kappa$ when there is only nearest-neighbor coupling between rods, of strength $U_1$. The matrix elements $T_\kappa(\theta', \theta)$ have the value $\exp(-\beta U_1)$, if the adjacent rods with orientations $\theta$ and $\theta'$ overlap, and 1 otherwise. The matrix has the obvious symmetries

$$T_\kappa(\theta, \theta') = T_\kappa(\pi - \theta, \pi - \theta'), \quad (1a)$$
$$T_\kappa(\theta, \theta') = T_\kappa(\pi - \theta', \pi - \theta). \quad (1b)$$

Therefore, it is sufficient to specify the matrix elements of $T_\kappa$ only for the range $\theta \in [0, \frac{\pi}{2}]$.

For $\kappa < \frac{1}{2}$, there are no overlaps. If $\frac{1}{2} < \kappa \leq \frac{1}{\sqrt{2}}$, an overlap of the nearest neighbor rods is possible, but only if $\cos\theta < \frac{1}{2\kappa}$ and $\theta' \in [\theta_{min}, \theta_{max}]$ (see figure 1a) where

$$\theta_{min}(\theta) = \pi - \arctan\left(\frac{\sin\theta}{\kappa^{-1} - \cos\theta}\right), \quad (2a)$$

$$\theta_{max}(\theta) = \pi + \theta - \arcsin\left(\frac{\sin\theta}{\kappa}\right). \quad (2b)$$

For $\frac{1}{\sqrt{2}} \leq \kappa \leq 1$, we get $T_\kappa(\theta', \theta) \neq 1$ if $\sin(\theta) \leq \kappa$, and $\theta' \in [\theta_{min}, \theta_{max}]$, where $\theta_{min}$ has different expressions for different ranges of the orientation $\theta$ of the right rod (see figure 1b). We get, for any $\theta$

$$\theta_{max} = \pi + \theta - \arcsin\left(\frac{\sin\theta}{\kappa}\right). \quad (3)$$

On the other hand, for $\theta_{min}$, we get, if $\theta \in [0, \arccos\left(\frac{1}{2\kappa}\right)]$, then

$$\theta_{min} = \pi - \arctan\left(\frac{\sin\theta}{\kappa^{-1} - \cos\theta}\right), \quad (4a)$$

whereas, if $\theta \in [\arccos\left(\frac{1}{2\kappa}\right), \arcsin(\kappa)]$, then we get

$$\theta_{min} = \arcsin\left(\frac{\sin\theta}{\kappa}\right) + \theta. \quad (4b)$$

For $\kappa > 1$, the elements $T_\kappa(\theta', \theta) \neq 1$ if $\theta' < \theta_{min}$ or $\theta' > \theta_{max}$, where $\theta_{max}$ has different expressions for different ranges of $\theta$ (see figure 1c). We get, for any $\theta$,

$$\theta_{min} = \theta - \arcsin\left(\frac{\sin\theta}{\kappa}\right). \quad (5)$$

On the other hand, if $\theta \in [0, \arccos\frac{1}{\kappa}]$, then

$$\theta_{max} = \arctan\left(\frac{\sin\theta}{\cos\theta - \kappa^{-1}}\right), \quad (6a)$$

if $\theta \in [\arccos\frac{1}{\kappa}, \arccos\frac{1}{2\kappa}]$, then

$$\theta_{max} = \pi + \arctan\left(\frac{\sin\theta}{\cos\theta - \kappa^{-1}}\right), \quad (6b)$$

and if $\theta \in [\arccos\frac{1}{2\kappa}, \frac{\pi}{2}]$, then

$$\theta_{max} = \theta + \arcsin\left(\frac{\sin\theta}{\kappa}\right). \quad (6c)$$

The shape of the boundary of the overlap regions $\theta_{max}$ and $\theta_{min}$, for different ranges of $\kappa$, is given in figure 2.

## II. EXACT DIAGONALIZATION FOR $\kappa = 1$

When $\kappa = 1$, the boundary of the overlap region is a set of straight lines and the associated transfer matrix $T_\kappa(\theta, \theta')$ is sketched in figure 3. This makes the calculation of the eigenvalue and associated eigenvector simple. The eigenequation is

$$\int_0^\pi \frac{d\theta'}{\pi} \psi_\kappa(\theta') T_\kappa(\theta', \theta) = \Lambda \, \psi_\kappa(\theta). \quad (7)$$

From (1b) we see that the eigenvector has the symmetry $\psi_\kappa(\theta) = \psi_\kappa(\pi - \theta)$. Considering this we write

$$\psi_\kappa(\theta) = \begin{cases} \psi_\kappa^{(1)}(\theta) & \text{for } 0 < \theta \leq \frac{\pi}{3}, \\ \psi_\kappa^{(2)}(\theta) & \text{for } \frac{\pi}{3} < \theta \leq \frac{\pi}{2}. \end{cases} \quad (8)$$

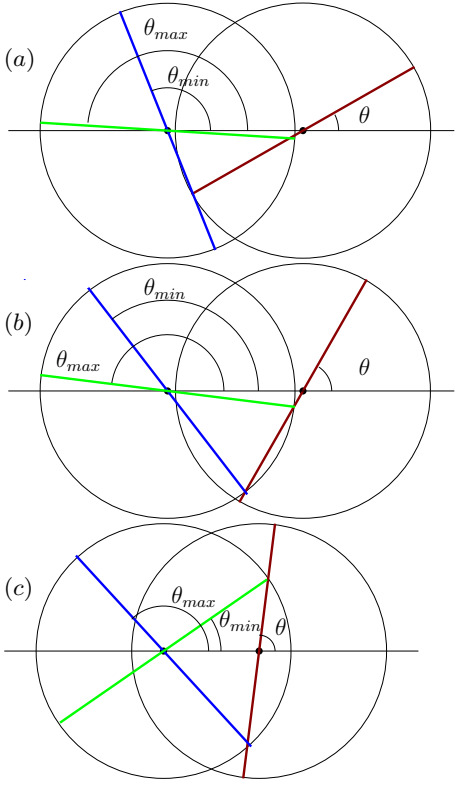

FIG. 1. Overlap criteria for a pair of nearest neighbor rods with orientation $(\theta', \theta)$ for values of $\kappa$ in the range (a) $\frac{1}{2} \leq \kappa < \frac{1}{\sqrt{2}}$, (b) $\frac{1}{\sqrt{2}} < \kappa < 1$, and (c) $\kappa > 1$. Overlap occurs for angle $\theta' \in [\theta_{min}, \theta_{max}]$, except in the last case where overlap is for angles $\theta' \in [0, \theta_{min}]$ or $\theta' \in [\theta_{max}, \pi]$.

Further, we define

$$\mathcal{N} = \int_0^{\frac{\pi}{2}} \frac{d\theta'}{\pi} \psi_\kappa(\theta').$$

Now, if we define $\psi_\kappa^{(1)}(\theta = \frac{\pi x}{3}) = P(x)$, and $\psi_\kappa^{(2)}(\theta = \frac{\pi}{2} - \frac{x\pi}{6}) = Q(x)$ for $0 \leq x \leq 1$, then the eigenvalue equation becomes

$$\mathcal{N} + \frac{1}{6}\int_0^x dy Q(y) + e^{-\beta U_1}\frac{1}{6}\int_x^1 dy Q(y)$$
$$+ e^{-\beta U_1}\frac{1}{3}\int_0^1 dy P(y) = \Lambda P(x), \quad (9a)$$

$$\mathcal{N} + \frac{1}{6}\int_0^1 dy Q(y) + \frac{1}{3}\int_x^1 dy P(y)$$
$$+ e^{-\beta U_1}\frac{1}{3}\int_0^x dy P(y) = \Lambda Q(x), \quad (9b)$$

with

$$\mathcal{N} = \frac{1}{3}\int_0^1 dx P(x) + \frac{1}{6}\int_0^1 dx Q(x). \quad (9c)$$

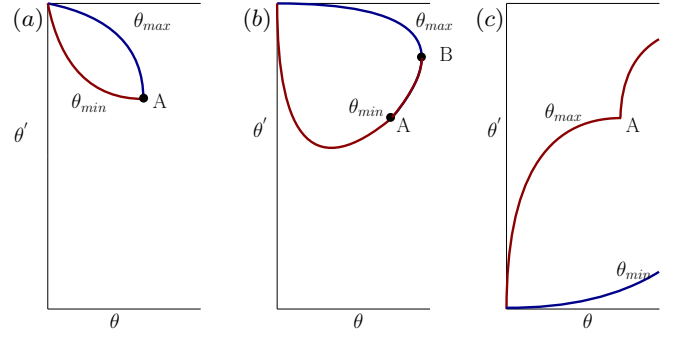

FIG. 2. The shape of the boundary of the overlap region in the transfer matrix for values of $\kappa$ in the range (a) $\frac{1}{2} \leq \kappa < \frac{1}{\sqrt{2}}$, (b) $\frac{1}{\sqrt{2}} < \kappa < 1$, and (c) $\kappa > 1$. Only the range $\theta < \frac{\pi}{2}$ is shown; the rest of the region can be constructed using the symmetry (1a, 1b). The point $A$ denotes $\theta = \arccos \frac{1}{2\kappa}$ and $B$ denotes $\theta = \arcsin \kappa$.

These integral equations can be converted into the following coupled differential equations

$$\Lambda \frac{dP}{dx} = \left(1 - e^{-\beta U_1}\right) \frac{Q(x)}{6},$$
$$\Lambda \frac{dQ}{dx} = -\left(1 - e^{-\beta U_1}\right) \frac{P(x)}{3}.$$

Solutions of these equations are given by

$$P(x) = \frac{N}{\Lambda}\left((1 + e^{-\beta U_1})\cos kx + \sqrt{2}\sin kx\right), \quad (10a)$$

$$Q(x) = \frac{N}{\Lambda}\left(2\cos kx - \sqrt{2}(1 + e^{-\beta U_1})\sin kx\right), \quad (10b)$$

where

$$\Lambda = \left(1 - e^{-\beta U_1}\right)/3\sqrt{2}k, \quad (10c)$$

$$k = \arctan\left(\frac{1 - e^{-\beta U_1}}{\sqrt{2}(2 + e^{-\beta U_1})}\right). \quad (10d)$$

Note that there is an infinite spectrum of eigenvalues. Other eigenvectors, and eigenvalues, including the antisymmetric ones can also be determined similarly.

### III. ANALYSIS OF THE SINGULARITIES

#### A. Singularity near $\kappa = \frac{1}{2}$

When $\kappa \leq \frac{1}{2}$, the elements of the transfer matrix

$$T_\kappa(\theta', \theta) = 1 \quad \text{for all } \theta \text{ and } \theta'. \quad (11)$$

The largest eigenvalue for this matrix is $\Lambda = 1$ and the corresponding eigenvector $\psi_{\frac{1}{2}}(\theta) = 1$. All other eigenvalues are zero. From a first order perturbation theory,

where $\kappa$ is varied around $\kappa = \frac{1}{2}$, the corresponding change in the largest eigenvalue is given by

$$\Delta\Lambda = \int_0^\pi \frac{d\theta}{\pi} \int_0^\pi \frac{d\theta'}{\pi} \psi_{\frac{1}{2}}(\theta) \Delta T(\theta,\theta') \psi_{\frac{1}{2}}(\theta')$$
$$= \langle \psi_{\frac{1}{2}} | \Delta T | \psi_{\frac{1}{2}} \rangle, \qquad (12)$$

where $\Delta T$ denotes the corresponding change in the transfer matrix. If we write, $\kappa = \frac{1}{2} - \varepsilon$, with positive and small $\varepsilon$, the transfer matrix will remain the same, i.e. $\Delta T = 0$, and therefore $\Delta\Lambda = 0$. However, if we write $\kappa = \frac{1}{2} + \varepsilon$, with positive and small $\varepsilon$, the corresponding change in the eigenvalue is

$$\Delta\Lambda = \left(e^{-\beta U_1} - 1\right) \Delta A, \qquad (13a)$$

where $\Delta A = A(\kappa = \frac{1}{2} + \varepsilon) - A(\kappa = \frac{1}{2})$ is the change in the overlap region, which is given by

$$\Delta A = 2 \int_0^{\arccos \frac{1}{2\kappa}} \frac{d\theta}{\pi} \int_{\theta_{min}(\theta)}^{\theta_{max}(\theta)} \frac{d\theta'}{\pi}. \qquad (13b)$$

To evaluate this change for $\varepsilon \to 0$ we see

$$\arccos \frac{1}{2\kappa} \simeq 2\sqrt{\varepsilon},$$

and consequently

$$\theta_{min} \simeq \pi - \theta(1 + 4\varepsilon) + \theta^3,$$
$$\theta_{max} \simeq \pi - \theta(1 - 4\varepsilon) - \theta^3.$$

Using these in the equation (13a, 13b), we get

$$\Delta\Lambda = \frac{16\varepsilon^2}{\pi^2} \left(e^{-\beta U_1} - 1\right). \qquad (14)$$

Then, the free energy per site is given by

$$\mathcal{F}(\kappa,\beta) = \begin{cases} 0 & \text{for } \kappa = \frac{1}{2} - \varepsilon, \\ -\frac{16\varepsilon^2}{\pi^2}\left(e^{-\beta U_1} - 1\right) & \text{for } \kappa = \frac{1}{2} + \varepsilon. \end{cases} \qquad (15)$$

Therefore, at $\kappa = \frac{1}{2}$ the free energy has a discontinuous second derivative.

### B. Singularity near $\kappa = 1$

The structure of the excluded region for $\kappa = 1 - \varepsilon$, with $\varepsilon > 0$, is shown in figure 4. We show here that the shaded area in this plot varies as $\varepsilon \log \frac{1}{\varepsilon}$.

As the transfer matrix has the symmetry (1a, 1b), the change in area $\Delta A$ is four times the area of the shaded region in figure 4. One of its boundary is a straight line $\theta' = \frac{1}{2}\theta + \frac{\pi}{2}$. To show that asymptotic shape of this boundary near $(0, \pi/2)$ is a hyperbola, we introduce the re-scaled coordinates $(\xi, \eta)$ using

$$\xi = \frac{\theta}{\sqrt{\varepsilon}}, \qquad \eta = \frac{\theta_{min} - \left(\frac{\pi}{2} + \frac{\theta}{2}\right)}{\sqrt{\varepsilon}}. \qquad (16)$$

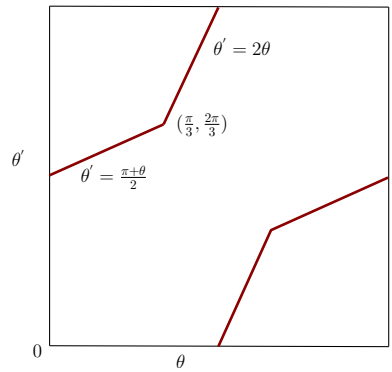

FIG. 3. The transfer matrix at $\kappa = 1$, with straight line boundaries of the overlap region.

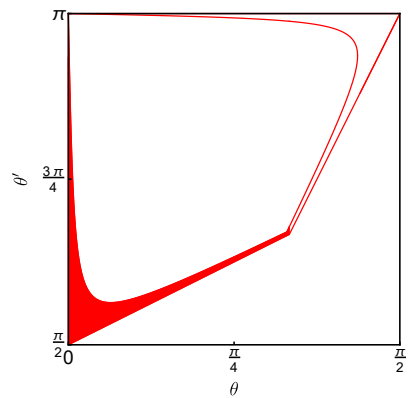

FIG. 4. The shaded area is one fourth of the change in area $\Delta A$ in the transfer matrix as $\kappa$ is decreased from 1 by an amount 0.02 (see figure 6 in the *Letter*).

Writing the equation (4a) in terms of this scaled coordinates, and solving in the limit $\varepsilon \to 0$, we get a scaled hyperbolic curve $\eta \xi = 1$. This implies, that to the leading order in small $\varepsilon$, the curved boundary of the shaded region in figure 4 follows $\eta = \frac{1}{\xi}$. Then, the area of this shaded region is $\varepsilon \int \eta \, d\xi$, where the upper limit of the integral varies as $\frac{1}{\sqrt{\varepsilon}}$. Therefore, we find that the area varies as $\varepsilon \log \frac{1}{\varepsilon}$. Keeping the exact pre-factors in our calculation, we get for small $\varepsilon$,

$$\Delta A \simeq 4\left(\varepsilon \ln \frac{\pi^2}{6} - \varepsilon \ln \varepsilon + \varepsilon\right). \qquad (17)$$

## IV. PROBABILITY DISTRIBUTION OF THE ORIENTATIONS OF A ROD

The probability distribution of the orientations of a rod shows a complex dependence on the angle $\theta$ and the parameter $\kappa$. The results of a simulation on a system of 100 rods and averaged over $10^6$ sample configurations is shown in figure 5. For $\kappa \leq \frac{1}{2}$, where the rods do not interact, the distribution is uniform. As $\kappa$ is increased

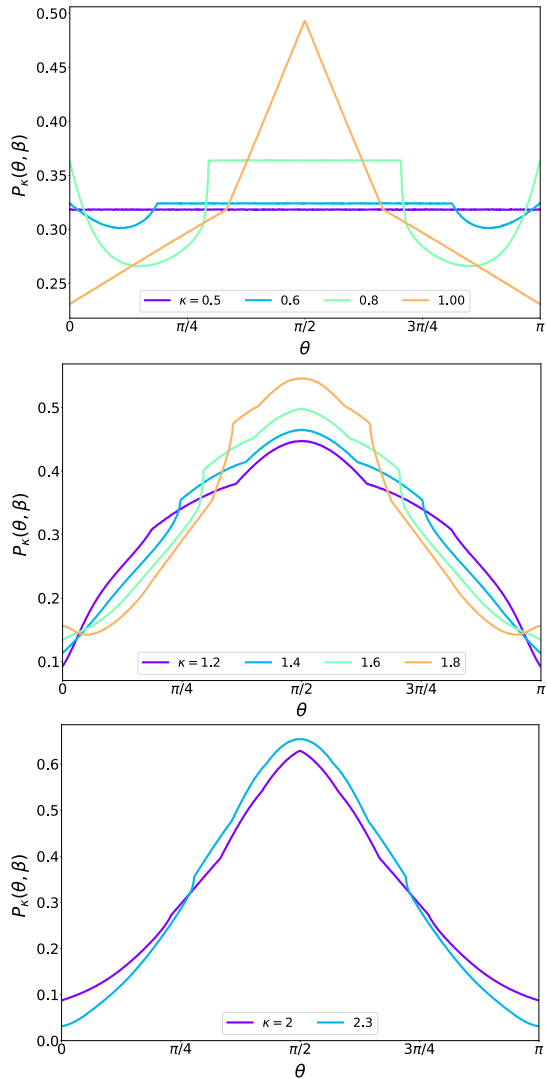

FIG. 5. Probability distribution of orientation of a rod measured in a Monte Carlo simulation of $N = 100$ rods with $\beta U_i = 1$ for all $i \geq 1$. The curves corresponds to different values of $\kappa$ indicated at the bottom panel of each figure.

above $\frac{1}{2}$, the distribution function shows a discontinuous first derivative when $\cos\theta = \frac{1}{2\kappa}$. This derivative discontinuity becomes a cusp singularity for $\kappa > \frac{1}{\sqrt{2}}$. The two symmetrically located cusps move in position with increasing values of $\kappa$ and merge at $\kappa = 1$. At this value, a new pair of singularities develop, and for the entire range $1 < \kappa < 2$, there are in total four singularities in the distribution function. At $\kappa = 2$, two of these singularities merge, but an additional pair of singularities emerge. This can be observed in figure 5. We find that whenever $\kappa$ crosses an integer multiple of $\frac{1}{\sqrt{2}}$, a new pair of cusp singularities develop, and then move towards each other as $\kappa$ is varied.

In the main text, we discussed the characterization in terms of the fractional number of $k$-th neighbors that

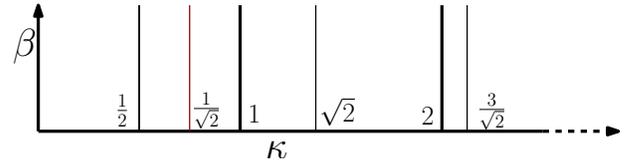

FIG. 6. A schematic phase diagram showing the different phase transition lines in the $\kappa$-$\beta$ plane, when all $U_i$ are equal. All lines are parallel to the $\beta$-axis.

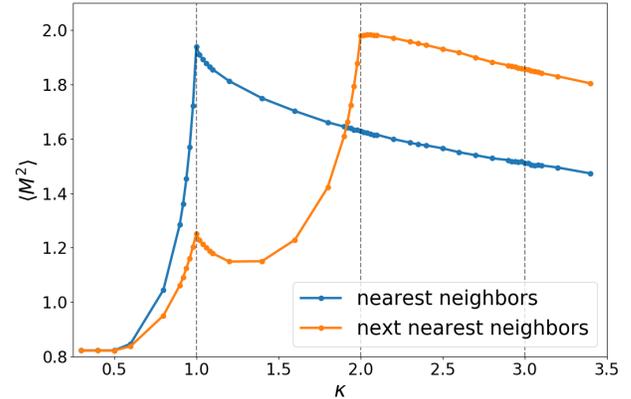

FIG. 7. Variance of the angle $\langle M^2 \rangle = \frac{1}{L}\langle \left[\sum_i (\theta_i - \frac{\pi}{2})\right]^2 \rangle$ for a system of rods with only nearest neighbor interactions and another system with upto next nearest neighbor interaction.

overlap directly. We can consider a finer characterization of the phases, by also using the number of cusps in the orientation distribution. If we do this, then one gets phase transitions whenever $\kappa$ is an integer multiple of 1 or $\frac{1}{\sqrt{2}}$. A schematic of such a phase diagram is drawn in figure 6.

The singularities can also be seen in the fluctuations. In figure 7, we show the numerical result for the variance of the orientations of a rod. Here we compare two cases: one with only $U_1$ non-zero, and the second with only $U_1$ and $U_2$ non-zero. In the first case, the singularity appears only at $\kappa = 1$, while in the second case, there is an additional singularity at $\kappa = 2$, but no detectable singularity at $\kappa = 3$.


	\end{center}
\end{widetext}

\end{document}